\documentclass[aps,preprint,superscriptaddress,showpacs,showkeys]{revtex4}
\usepackage[dvipdf]{graphicx}
\usepackage{amssymb}
\usepackage{amsbsy}
\usepackage{latexsym}
\usepackage{epsfig}
\usepackage{mathdots}

\newcommand{\be}{\begin{equation}}
\newcommand{\ee}{\end{equation}}
\newcommand{\ba}{\begin{eqnarray}}
\newcommand{\ea}{\end{eqnarray}}

\newcommand{\dg}{\dagger}
\newcommand{\wbe}{\begin{widetext}}
\newcommand{\wee}{\end{widetext}}

\begin{document}

\title{Quantization of  $n$ coupled  scalar field theory}

\author{Yong-Wan Kim}

\email{ywkim65@gmail.com}

\affiliation {Center for Quantum Spacetime, Sogang University, Seoul
121-742, Korea}

\author{Yun Soo Myung}

\email{ysmyung@inje.ac.kr}

\affiliation {Institute of Basic Science and School of Computer
Aided Science, Inje University, Gimhae 621-749, Korea}

\author{Young-Jai Park}

\email{yjpark@sogang.ac.kr}

\affiliation {Center for Quantum Spacetime, Sogang University, Seoul
121-742, Korea}

\affiliation {Department of Physics, Sogang University, Seoul
121-742, Korea}

\begin{abstract}
We study a model of $n$ coupled  scalar fields  in Minkowski
spacetime where all masses degenerate, which is considered as a toy
model of polycritical gravity on AdS spacetime. We quantize this
model within the Becchi-Rouet-Stora-Tyutin (BRST) scheme by
introducing $n$ Faddeev-Popov (FP) ghost fields. Extending a BRST
quartet generated by two scalars and two FP ghosts to $n$ scalars
and $n$ FP ghosts, there remains a physical subspace with positive
norm for odd $n$, but there exists only the vacuum for even $n$.
This clearly shows a non-triviality of odd-higher order derivative
scalar field  theories.  This is helpful to understand  the
truncation mechanism which is used to obtain a unitary conformal
field theory dual to  linearized polycritical gravity. It turns out
that the truncation mechanism is nothing but a general quartet
mechanism appeared when introducing the FP ghost action.

\end{abstract}



\maketitle

\section{Introduction}

The quantization of the system with first-class
constraints~\cite{Fradkin:1975cq} has been performed using the BRST
symmetry~\cite{Becchi:1975nq,Tyutin:1975qk,Kugo:1979gm}. The system
with second-class constraints could  be quantized by converting
these to a first-class theory in an extended phase
space~\cite{Kim:1992ey,Restuccia:1992gs}.

Even though we are seeking to find a consistent quantum
gravity~\cite{Stelle:1976gc,Barth:1983hb}, we
 focus on the quantization of the scalar theory but not the gauge and
gravity theories because of its simplicity.  To this end, a chiral
boson is a well-known example of the second-class theory in two
dimensions.  After the BRST quantization of a chiral boson, the
quartet mechanism forces all states to have a zero norm, leaving the
vacuum~\cite{Girotti:1992hy}.  Importantly, it was argued that all
higher derivative scalar theories are trivial because these have the
BRST symmetry when introducing FP ghosts, and have only  the vacuum
when  imposing  the  BRST quartet~\cite{Rivelles:2003jd}. However,
for the $2n$-order Klein-Gordon theory with different
masses~\cite{Barth:1983hb}, the odd $n$  and the even $n$ cases
feature qualitative differences. For  odd $n$, one has $(n-1)/2$
ghost (physical) fields and $(n+1)/2$ physical (ghost) fields
according to the overall negative (positive) sign of the free part
of the Lagrangian~\cite{de Urries:1998bi}. Here ghost (physical)
fields represent their scalar propagators  with negative (positive)
norm states.  For  even $n$, one finds $n/2$ fields of each type.
This distinguishes the odd $n$ case  from the even $n$ case.
However, the degenerate cases are ruled out in that approach because
the higher-order Green's function (equivalent second-order
Lagrangian) blows up after performing the partial fraction.

In this work, we investigate a model of $n$ coupled  scalar fields
in Minkowski spacetime where all masses degenerate, which leads to a
$2n$-order single scalar theory when  eliminating $n-1$ auxiliary
scalar fields~\cite{Rivelles:2003jd}. In particular, for $n=2$, the
model was  considered as a toy model of the fourth-order critical
gravity  on AdS$_3$ spacetime which appeared in the pursuit of
quantum gravity. Both of $n$ coupled scalar field and polycritical
gravity theories may have the same rank-$n$  logarithmic conformal
field theory (LCFT) as their
duals~\cite{Bergshoeff:2012sc,Moon:2012wr,Grumiller:2013at}, which
still suffers from the non-unitarity. A truncation mechanism has
been introduced to cure the
non-unitarity~\cite{Bergshoeff:2012ev,Kleinschmidt:2012rs}. However,
up to now, there is no consistent truncation mechanism to provide a
unitary CFT.  Furthermore, it was pointed out  that these linearized
approaches of polycritical gravities have pathologies when
considering the non-linear level~\cite{Apolo:2012vv}. This implies
that calculations on the linearized level seemed to lend support to
the possibility of truncating the theory. In this sense, we have to
regard our model of the $n$-coupled scalar field theory as a toy
model of (linearized) polycritical gravities.

We wish to quantize the $n$ coupled scalar field theory  within the
BRST quantization scheme by constructing the FP ghost action
composed of $n$ FP ghost fields. Extending a BRST quartet generated
by two scalars and FP ghosts to $n$ scalars and FP ghosts, there
remains a physical subspace with positive norm for odd $n$, while
there exists only the vacuum for even $n$. This  shows the
non-triviality of odd-higher order derivative scalar field theories
clearly, which might provide a hint to resolve the non-unitarity
issue appeared in developing the higher-derivative quantum gravity.

\section{ $n$ coupled  scalar field theory}

Let us start with  the $n$  coupled scalar field model with
degenerate masses~\cite{Bergshoeff:2012sc}
 \ba
 \label{action}
 S_0&=&-\frac{1}{2}\int d^4x
 \sum_{i,j=1}^n(X_{ij}\partial_\mu\phi_i\partial^\mu\phi_j+Y_{ij}\phi_i\phi_j),
 \ea
where we adopt the Minkowskian  convention of  $\eta_{\mu\nu}={\rm
diag.}(-++\,+)$ and $x^\mu=(t,\vec{x})$.  The $(n\times n)$-matrices
$X_{ij}$ and $Y_{ij}$ for the scalar fields are given by
 \ba
 X_{ij}
 =\left(
  \begin{array}{ccccc}
   0      & \cdots  &0       & 0 & 1 \\
   \vdots &         &0       & 1 & 0 \\
   0      & \iddots &\iddots &   & \vdots \\
   0      & 1       &        &   &   \\
   1      & 0       &\cdots  &   & 0 \\
  \end{array}
 \right),~~
  Y_{ij}
 =\left(
  \begin{array}{ccccc}
   0      & \cdots  & 0 & 1      & m^2 \\
   \vdots &         & 1 & m^2    & 0 \\
   0      & \iddots & \iddots  & & \vdots \\
   1      & m^2     &  &  & \\
   m^2    & 0       & \cdots  & & 0 \\
  \end{array}
 \right)
 \ea
for $i,~j=1,...,n$ ($n\ge 2$). In this work we will not consider the
non-degenerate case with different masses for  $Y_{ij}$, because it
could not be considered as a toy model of the polycritical gravity.

 The equations of motion for the $n$
coupled scalar fields are
 \ba\label{eomscal}
 && (\square-m^2)\phi_1=0,\\
 \label{eomscal2}&& (\square-m^2)\phi_i=\phi_{i-1},~~~i=2,... ,n,
 \ea
which lead to
 \be
  (\square-m^2)^n\phi_n=0
 \ee
for a scalar field $\phi_n$. In the classical aspect of the theory,
the other fields $\phi_l$ with $l=1, \cdots , n-1$ are considered as
auxiliary fields used to lower the number of derivatives in the
single scalar $\phi_n$ action~\cite{Rivelles:2003jd} \be
\label{saction} S_n=-\frac{1}{2}\int d^4x
(\square-m^2)^{\frac{n}{2}}\phi_n (\square-m^2)^{\frac{n}{2}}\phi_n.
\ee However, in the quantum aspect of the theory,
$\{\phi_n,\phi_l\}$ will be treated equally as scalar fields.

In order to obtain the BRST invariant action, we have to construct
the corresponding FP ghost action. Usually, the BRST symmetry was
found in gauge theories as a symmetry of the gauge-fixed
action~\cite{Becchi:1975nq,Tyutin:1975qk,Kugo:1979gm}. Its purpose
is definitely to remove unphysical fields (negative norm states)
associated with gauge invariance. On the other hand, physical
fields  are defined as those which have zero ghost number and are
invariant under BRST transformations.

The BRST symmetry in this work  is not due to gauge symmetry after a
gauge-fixing. Surely, it takes into account a feature of giving the
higher-order derivative structure starting from the second-order
action (\ref{action}) via the scalar coupling.  At this stage, we
emphasize that the model (\ref{action}) [or (\ref{saction})]
inherently possesses ghost states.  In order to eliminate the ghosts
arising from the higher derivative action, we need to construct the
corresponding FP ghost action.  Here, we use the same FP terminology
which was used in the gauge and gravity theories to distinguish
between FP ghosts  and  ghost fields with negative norm state
(poltergeist~\cite{Barth:1983hb}). In the U(1) gauge theory, the FP
ghosts are introduced to remove the unphysical fields of scalar and
longitudinal photons by imposing the quartet, leaving two transverse
photons~\cite{Kugo:1979gm}. More precisely, the `gauge' FP ghosts
are used for the quantization of gauge and gravity theories, while
the `higher-derivative' FP ghosts are introduced to take into
account the higher-derivative nature of the $n$ coupled scalar field
theory (\ref{action}) [or (\ref{saction})].

Recently, we have studied a sixth order derivative ($n=3$) scalar
field model in Minkowski spacetime in a BRST invariant
manner~\cite{Kim:2013waf} as a toy model of critical gravity
theories. There, the `higher-derivative' FP ghost action  was
included to require that the resultant action is invariant under the
BRST transformation.   By extending the $n=3$ analysis to the $n$
coupled  scalar field theory,  we find the ghost action composed of
$n$ FP ghost fields $c_i$ as
  \ba
 \label{g-action}
 S_g&=&-\frac{1}{2}\int d^4x \sum_{i,j=1}^n(Z_{ij}\partial_\mu c_i\partial^\mu c_j+W_{ij}c_i c_j),
 \ea
whose even  $ Z_{ij}$ and  $W_{ij}$  are given by
 \ba\label{evenm}
 Z_{ij}
 =\left(
  \begin{array}{cccccc}
   0      &         &\cdots   &  0   &   0   & 1      \\
   0      &         &\cdots   &  0   &   1   & 0      \\
   0      &         &         &      &       & 0      \\
  \vdots  & \iddots &\iddots  &      &       & \vdots \\
   0      & -1      &\iddots  &      &       & 0      \\
   -1     &  0      &\cdots   &      &       & 0     \\
  \end{array}
 \right),~~
  W_{ij}
 =\left(
  \begin{array}{cccccccccc}
   0      &         &         &       &   &\cdots &         & 0      & 0     & m^2    \\
   0      &         &         &       &   &\cdots &         & 0      & m^2   & 1      \\
   0      &         &         &       &   &       &         &\iddots &\iddots& 0      \\
  \vdots  &         &         &       &   &       &         & 1      &\iddots& 0      \\
   0      &         &         &       &   & 0     &         &        &       & \vdots \\
   0      & \iddots & \iddots & -1    &   &       &         &        &       &        \\
   0      & -m^2    & \iddots &\iddots&   &       &         &        &       &        \\
   -m^2   & -1      &  0      &\cdots &   &       &         &        &       & 0      \\
  \end{array}
 \right),
 \ea
for $i,~j=1,...,2k(=n)$. On the other hand,   odd
 $ Z_{ij}$ and  $W_{ij}$ matrices take the forms
 \ba\label{oddm}
 Z_{ij}
 =\left(
  \begin{array}{cccccc|c}
   0      &         & \cdots  &        0  & 0    & 1     & 0 \\
   0      &         & \cdots  &        0  & 1    & 0     & 0 \\
   0      &         &         &    ~~~    &      & 0     & 0 \\
   \vdots & \iddots & \iddots &           &      &\vdots &  \vdots      \\
   0      & -1      & \iddots &           &      &       &  \\
   -1     & 0       & \cdots  &           &      &       &    \\ \hline
   0      & 0       & \cdots  &           &      &       & 0 \\
  \end{array}
 \right),~~
 W_{ij}
 =\left(
 \begin{array}{cccccccccc|c}
   0      &         &         &       &   &\cdots&         & 0      & 0     & m^2    & 0\\
   0      &         &         &       &   &\cdots&         & 0      & m^2   & 1      & 0\\
   0      &         &         &       &   &      &         &\iddots &\iddots& 0      & 0\\
  \vdots  &         &         &       &   &      &         & 1      &\iddots& \vdots & \vdots\\
   0      &         &         &       &   & 0    &         &        &       & 0      & 0\\
   0      & \iddots & \iddots & -1    &   &      &         &        &       & 0      & 0\\
   0      & -m^2    & \iddots &\iddots&   &      &         &        &       &\vdots  &\vdots\\
   -m^2   & -1      &  0      &\cdots &   &      &         &        &       &        &\\ \hline
   0      &  0      &  0      &\cdots &   &      &         &        &       &        & 0

  \end{array}
 \right),
 \ea
for $i,~j=1,...,2k+1(=n)$.  We note that in the odd  case
(\ref{oddm}), the last-null row and column are added to the even
case (\ref{evenm}).  For non-degenerate case with different masses
for  $W_{ij}$ in (\ref{evenm}) and (\ref{oddm}), we could not find
the BRST invariant action because the BRST symmetry is not
nilpotent. We explain how the ghost action (\ref{g-action}) is
nontrivially constructed, depending on $n$ because the $n$ coupled
scalar action (\ref{action}) has already known. For $n=2,3$, two FP
ghosts is enough to have the BRST invariant action for  a dipole
ghost field (singleton). This was a known case. We need to introduce
more FP ghost fields to construct the BRST invariant action as $n$
increases. For example, we have 4 FP ghosts for $n=4,5$, 6 FP ghosts
for $n=6,7$, and so on. This indicates  clearly  nontrivial terms
for $n>3$ when comparing the known cases of $n\le 3$.

Now,  we show that the total action
 \be
 \label{actiont} S_t=S_0+S_g,
 \ee
is invariant under the  BRST transformation
 \ba \label{e-brst}
 && \delta\phi_1=0,~\cdots,~\delta\phi_k=0,~\delta\phi_{k+1}=c_k,~\cdots,~\delta\phi_{2k}=c_1, \nonumber\\
 && \delta c_1=0,~\cdots,~\delta c_k=0,~\delta c_{k+1}=\phi_k,~\cdots,~\delta c_{2k}=\phi_1,
 \ea
for the even ($n=2k$) case, while
 \ba \label{o-brst}
 && \delta\phi_1=0,~\cdots,~\delta\phi_k=0,~\underline{\delta\phi_{k+1}=0},~\delta\phi_{k+2}=c_k,~\cdots,~\delta\phi_{2k+1}=c_1, \nonumber\\
 && \delta c_1=0,~\cdots,~\delta c_k=0,~\delta c_{k+1}=\phi_k,~\cdots,~\delta c_{2k+1}=\phi_1,
 \ea
for the odd ($n=2k+1$) case. Here we wish to point out that for the
odd case of (\ref{o-brst}), there exists an additional
BRST-invariant field like $\phi_{k+1}$ when comparing it with the
even case of (\ref{e-brst}). In the odd $n$ coupled scalar theory,
only $\phi_{k+1}$ is a physical field, whereas all remaining fields
belong to unphysical fields.  This might explain  an origin of
existing a physical state with positive norm state.

Finally, we derive $n$ coupled equations for  $n$ ghost fields
 \ba
 &&\label{eomgh1} (\square-m^2)c_{i-1}=c_i,~~~i=2,... , k,\\
 &&\label{eomgh2} (\square-m^2)c_k=0,\\
 &&\label{eomgh3} (\square-m^2)c_{i-1}=c_i,~~~i=k+2,..., 2k,\\
 &&\label{eomgh4} (\square-m^2)c_{2k}=0,
 \ea
where $k=[\frac{n}{2}]$ ($n\ge 2$) is the greatest integer  which is
less  than $\frac{n}{2}$. Evidently, these are different from the FP
ghost equations of the gauge theory. Note that by making successive
eliminations of the smaller indices, these equations reduce to two
FP ghosts equations
 \ba
 && (\square-m^2)^k c_1=0,\\
 && (\square-m^2)^k c_{k+1}=0,
 \ea
which are just two FP ghost equations for a single field $\phi_n$
(\ref{saction})~\cite{Rivelles:2003jd}.

\section{BRST transformations of modes}

In this section, instead of scalar and FP ghost fields
$\{\phi_i(x),c_i(x)\}$, we find the BRST transformations of
corresponding modes from solutions to Eqs.
(\ref{eomscal})-(\ref{eomscal2}) and
(\ref{eomgh1})-(\ref{eomgh4}). First of all, making use of an
ansatz
 \be
 \phi_1(x)=\int\frac{d^3k}{(2\pi)^{3/2}\sqrt{2\omega}}\phi_{1}(\vec{k},t)e^{i\vec{k}\cdot\vec{x}},
 \ee
Eq. (\ref{eomscal}) becomes one dimensional equation for $
\phi_{1}(\vec{k},t)$ as
 \be\label{eomf}
 \left(\frac{d^2}{dt^2}+\omega^2\right)\phi_{1}(\vec{k},t)=0
 \ee
with $\omega^2=\vec{k}^2+m^2$. This is solved to give  a solution
 \be\label{fsol}
 \phi_{1}(\vec{k},t)= iN_1
       \left(a_1(\vec{k})e^{-i\omega t}-a^\dg_1(\vec{k})e^{i\omega t}\right)
 \ee
with two Fourier  modes $a_1(\vec{k})$ and $a^\dg_1(\vec{k})$. Here,
we have introduced a coefficient $N_1$ which may be taken to be 1.
On the other hand, choosing $N_1=-\sqrt{m}$  reproduces the particle
theory's result appeared in Ref.~\cite{Rivelles:2003jd}. Using the
ansatz for Eq. (\ref{eomscal2})
 \be\label{solfind1}
 \phi_i(x)=\int\frac{d^3k}{(2\pi)^{3/2}\sqrt{2\omega}}\phi_{i}(\vec{k},t)e^{i\vec{k}\cdot\vec{x}},
 \ee
we  obtain the time-dependent equations as
 \be\label{eomf1}
 \left(\frac{d^2}{dt^2}+\omega^2\right)\phi_{i}(\vec{k},t)=-\phi_{i-1}(\vec{k},t),~~~i=2,...,n.
 \ee
Eq. (\ref{eomf1}) can be further separated into two first-order
differential equations as
 \ba
 &&\left(\frac{d}{dt}+i\omega\right)\psi_i(\vec{k},t)=-\phi_{i-1}(\vec{k},t),\\
 &&\left(\frac{d}{dt}-i\omega\right)\phi_i(\vec{k},t)=\psi_i(\vec{k},t),
 \ea
whose solutions are  given by
 \ba
 &&\psi_i(\vec{k},t)=-e^{-i\omega t}\int dt' e^{i\omega t'}\phi_{(i-1)}(\vec{k},t'),\\
 &&\label{solfind2}
 \phi_i(\vec{k},t)=e^{i\omega t}\int dt' e^{-i\omega t'}\psi_i(\vec{k},t'),
 \ea
respectively.

As a result, we find the first-five iterative solutions as
 \ba
 \label{sol1}
 \phi_1(x)&=&\int\frac{d^3k}{(2\pi)^{3/2}\sqrt{2\omega}} (iN_1)
       \left(a_1(\vec{k})e^{-i\omega t+i\vec{k}\cdot\vec{x}}-{\rm c.c}\right), \\
 \label{sol2} \phi_2(x)&=&\int\frac{d^3k}{(2\pi)^{3/2}\sqrt{2\omega}}\left(-\frac{N_1}{2\omega^2}\right)
        \left[\left(a_2(\vec{k})-a_1(\vec{k})\omega t\right)e^{-i\omega t+i\vec{k}\cdot\vec{x}}
                    +{\rm c.c}\right],\\
 \label{sol3} \phi_3(x)&=&\int\frac{d^3k}{(2\pi)^{3/2}\sqrt{2\omega}}\left(-\frac{iN_1}{4\omega^4}\right)
          \left[\left(a_3(\vec{k})-\left(\frac{i}{2}a_1(\vec{k})\omega t+a_2(\vec{k})\right)\omega t
                      +\frac{1}{2}a_1(\vec{k})\omega^2t^2\right)e^{-i\omega t+i\vec{k}\cdot\vec{x}}\right.\nonumber\\
     &&~~~~~~~~~~~~~~\left. - \frac{}{}{\rm c.c}\right],\\
 \label{sol4} \phi_4(x)&=&
   \int\frac{d^3k}{(2\pi)^{3/2}\sqrt{2\omega}}\left(\frac{N_1}{8\omega^6}\right) \left[(a_4(\vec{k})
               +\left(\frac{1}{2}a_1(\vec{k})-\frac{i}{2}a_2(\vec{k})-a_3(\vec{k})\right)\omega t\right.\nonumber\\
     &&~~~~~~~~~~~~~~\left.\left.  +\left(\frac{i}{2}a_1(\vec{k})+\frac{1}{2}a_2(\vec{k})\right)\omega^2t^2
                                 -\frac{1}{6}a_1(\vec{k})\omega^3t^3\right)e^{-i\omega t+i\vec{k}\cdot\vec{x}}+{\rm c.c}\right],\\
 \label{sol5}\phi_5(x)&=&
   \int\frac{d^3k}{(2\pi)^{3/2}\sqrt{2\omega}}\left(\frac{iN_1}{16\omega^8}\right)
             \left[\left(a_5(\vec{k})+\left(\frac{5i}{8}a_1(\vec{k})+\frac{1}{2}a_2(\vec{k})
                            -\frac{i}{2}a_3(\vec{k})-a_4(\vec{k})\right)\omega t\right.\right.  \nonumber\\
     &&~~~~~~~~~~~~~~+\left(-\frac{5}{8}a_1(\vec{k})+\frac{i}{2}a_2(\vec{k})+\frac{1}{2}a_3(\vec{k})\right)\omega^2t^2
                  -\left(\frac{i}{4}a_1(\vec{k})+\frac{1}{6}a_2(\vec{k})\right)\omega^3t^3\nonumber\\
     &&~~~~~~~~~~~~~~\left.\left. +\frac{1}{24}a_1(\vec{k})\omega^4t^4\right)e^{-i\omega t+i\vec{k}\cdot\vec{x}}+{\rm
     c.c}\right]
 \ea
with five sets of Fourier modes $\{a_i,a^{\dg}_i\}$. Here, we
observe that $\phi_i \sim t^{i-1}$ reflects the classical solution
to a higher-order degenerate equation of $(\square-m^2)^i\phi_i=0$
and an inclusion of all previous modes in $\phi_i$ represents the
coupled nature of the second-order equation
[$(\square-m^2)\phi_i=-\phi_{i-1}$] in Eq.~(\ref{eomscal2}).

At this stage,  it is appropriate to comment that we have
iteratively found  the solutions to  (\ref{eomscal2}) up to $n=7$
through the steps of (\ref{solfind1})-(\ref{solfind2}). However, we
encounter  a difficulty to  write them down in a compact way.

Similarly, introducing  the ansatz for the $n$ ghost fields
 \be
 c_i(x)=\int\frac{d^3k}{(2\pi)^{3/2}\sqrt{2\omega}}c_{i}(\vec{k},t)e^{i\vec{k}\cdot\vec{x}},
 \ee
Eqs. (\ref{eomgh1})-(\ref{eomgh4}) are reduced to one dimensional
equations
 \ba\label{eomghptl}
 && \left(\frac{d^2}{dt^2}+\omega^2\right)c_{i-1}(\vec{k},t)=-c_{i}(\vec{k},t),~~~i=2,... , k,\\
 && \left(\frac{d^2}{dt^2}+\omega^2\right)c_k(\vec{k},t)=0,\\
 && \left(\frac{d^2}{dt^2}+\omega^2\right)c_{i-1}(\vec{k},t)=-c_{i}(\vec{k},t),~~~i=k+2,..., 2k,\\
 && \left(\frac{d^2}{dt^2}+\omega^2\right)c_{2k}(\vec{k},t)=0.
 \ea
Corresponding to the solutions of the $n=5$ coupled  scalar  field theory, we write down the first-four ghost solutions.\\
For $n=2$ case,
 \ba
 c_1(x)&=&\int\frac{d^3k}{(2\pi)^{3/2}\sqrt{2\omega}}\left(-\frac{N_1}{2\omega^2}\right)
           \left(c_1(\vec{k})e^{-i\omega t+i\vec{k}\cdot\vec{x}}+{\rm c.c}\right),\\
 c_2(x)&=&\int\frac{d^3k}{(2\pi)^{3/2}\sqrt{2\omega}}\left(iN_1\right)
           \left(c_2(\vec{k})e^{-i\omega t+i\vec{k}\cdot\vec{x}}-{\rm c.c}\right).
 \ea
For $n=3$ case,
 \ba
 c_1(x)&=&\int\frac{d^3k}{(2\pi)^{3/2}\sqrt{2\omega}}\left(-\frac{iN_1}{4\omega^4}\right)
          \left(c_1(\vec{k})e^{-i\omega t+i\vec{k}\cdot\vec{x}}-{\rm c.c}\right),\\
  c_2(x)&=&\int\frac{d^3k}{(2\pi)^{3/2}\sqrt{2\omega}}\left(iN_1\right)
           \left(c_2(\vec{k})e^{-i\omega t+i\vec{k}\cdot\vec{x}}-{\rm c.c}\right).
 \ea
For $n=4$ case,
 \ba
 c_1(x)&=&\int\frac{d^3k}{(2\pi)^{3/2}\sqrt{2\omega}}\left(\frac{N_1}{8\omega^6}\right)
            \Big[\Big(c_1(\vec{k})-c_2(\vec{k})\omega t\Big)e^{-i\omega t+i\vec{k}\cdot\vec{x}}+{\rm c.c}\Big],\\
 c_2(t)&=&\int\frac{d^3k}{(2\pi)^{3/2}\sqrt{2\omega}}\left(-\frac{iN_1}{4\omega^4}\right)
          \left(c_2(\vec{k})e^{-i\omega t+i\vec{k}\cdot\vec{x}}-{\rm c.c}\right),\\
 c_3(t)&=&\int\frac{d^3k}{(2\pi)^{3/2}\sqrt{2\omega}}\left(-\frac{N_1}{2\omega^2}\right)
            \Big[\Big(c_3(\vec{k})-c_4(\vec{k})\omega t\Big)e^{-i\omega t+i\vec{k}\cdot\vec{x}}+{\rm c.c}\Big],\\
 c_4(t)&=&\int\frac{d^3k}{(2\pi)^{3/2}\sqrt{2\omega}}\left(iN_1\right)
           \left(c_4(\vec{k})e^{-i\omega t+i\vec{k}\cdot\vec{x}}-{\rm c.c}\right).
 \ea
Finally, for $n=5$ case,
 \ba
 c_1(x)&=&\int\frac{d^3k}{(2\pi)^{3/2}\sqrt{2\omega}}\left(\frac{iN_1}{16\omega^8}\right)
            \Big[\Big(c_1(\vec{k})-c_2(\vec{k})\omega t\Big)e^{-i\omega t+i\vec{k}\cdot\vec{x}}-{\rm c.c}\Big],\\
   c_2(x)&=&\int\frac{d^3k}{(2\pi)^{3/2}\sqrt{2\omega}}\left(-\frac{N_1}{8\omega^6}\right)
           \left(c_2(\vec{k})e^{-i\omega t+i\vec{k}\cdot\vec{x}}+{\rm c.c}\right),\\
 c_3(t)&=&\int\frac{d^3k}{(2\pi)^{3/2}\sqrt{2\omega}}\left(-\frac{N_1}{2\omega^2}\right)
            \Big[\Big(c_3(\vec{k})-c_4(\vec{k})\omega t\Big)e^{-i\omega t+i\vec{k}\cdot\vec{x}}+{\rm c.c}\Big],\\
 c_4(t)&=&\int\frac{d^3k}{(2\pi)^{3/2}\sqrt{2\omega}}\left(iN_1\right)
           \left(c_4(\vec{k})e^{-i\omega t+i\vec{k}\cdot\vec{x}}-{\rm c.c}\right).
 \ea
Here we distinguish number of ghost modes between 2 for  $n=2,3$ and
4 for $n=4,5$.   With these,  we obtain the BRST transformation for
all modes $a_i(\vec{k})$ and $c_i(\vec{k})$ as
 \ba \label{e-brstptl}
 && \delta a_1(\vec{k})=0,~\cdots,~\delta a_k(\vec{k})=0,
          ~\delta a_{k+1}(\vec{k})=c_k(\vec{k}),~\cdots,~\delta a_{2k}(\vec{k})=c_1(\vec{k}), \nonumber\\
 && \delta c_1(\vec{k})=0,~\cdots,~\delta c_k(\vec{k})=0,
          ~\delta c_{k+1}(\vec{k})=a_k(\vec{k}),~\cdots,~\delta c_{2k}(\vec{k})=a_1(\vec{k}),
 \ea
for the even ($n=2k$) case, while
 \ba \label{o-brstf}
 && \delta a_1(\vec{k})=0,~\cdots,~\delta a_k(\vec{k})=0,~\underline{\delta a_{k+1}(\vec{k})=0},
       ~\delta a_{k+2}(\vec{k})=c_k(\vec{k}),~\cdots,~\delta a_{2k+1}(\vec{k})=c_1(\vec{k}), \nonumber\\
 && \delta c_1(\vec{k})=0,~\cdots,~\delta c_k(\vec{k})=0,
       ~\delta c_{k+1}(\vec{k})=a_k(\vec{k}),~\cdots,~\delta c_{2k+1}(\vec{k})=a_1(\vec{k}),
 \ea
for the odd ($n=2k+1$) case.  Note from Eq.~(\ref{o-brstf}) that the
mode $a_{k+1}(\vec{k})$ is invariant under the BRST transformation.
In order to see the role of  remaining modes, we have to compute all
commutators between the modes.

\section{General quartet mechanism}

After a tedious  computation, we  derive the first-four commutation
relations  between $A_a$ and $A^\dagger_b$  where
$A_a~(A_b^\dagger)$ denotes the set of the modes
$\{a_i(\vec{k}),c_j(\vec{k})\}~(\{a_i^\dagger(\vec{k}),c_j^\dagger(\vec{k})\})$
with $i=1,\cdots,n$, $j=1,\cdots,n$ for even $n$ and $i=1,\cdots,n$,
$j=1,\cdots,n-1$ for odd $n$.

For $n=2$, one has
 \be
 [A_a, A^\dg_b]_\mp=
 \frac{2\omega^2}{N^2_1} \left(
  \begin{array}{c|cccc}
       &   a^\dg_2 & a^\dg_1 & c^\dg_1 & c^\dg_2 \\ \hline
   a_2 & 0 & i &  0  & 0 \\
   a_1 & -i  & -1 &  0  & 0 \\
   c_1 & 0  &  0  &  0  &  -i\\
   c_2 & 0  &  0  &  -i & 0 \\
  \end{array}
 \right)\delta^3(\vec{k}-\vec{k'}),
 \ee
which form a quartet to give the zero norm state.  This was designed
for a dipole ghost pair for the
singleton~\cite{Kugo:1979gm,Rivelles:2003jd,Myung:1999nd}. Note that
the subscripts $-~(+)$ denote the commutator (anti-commutator) for
the bosonic (fermionic) fields. On the other hand, commutators
between bosonic and fermionic fields vanish.

For $n=3$ in Ref.~\cite{Kim:2013waf}, the commutators take the
forms
 \be
 [A_a, A^\dg_b]_\mp=
 \frac{4\omega^4}{N^2_1}\left(
  \begin{array}{c|ccccc}
      &  a^\dg_2 &  a^\dg_1 & a^\dg_3 & c^\dg_1 & c^\dg_2 \\ \hline
  a_2 & 1 & 0& -i & 0 & 0 \\
  a_1 &  0 & 0 & -1 & 0 & 0 \\
  a_3 & i & -1 & \frac{3}{2} & 0 & 0   \\
  c_1 &  0 & 0 & 0 &  0  &  -1\\
  c_2 &  0 & 0 & 0 & 1 & 0 \\
  \end{array}
 \right)\delta^3(\vec{k}-\vec{k'}),
 \ee
which shows that
$[a_2(\vec{k}),a_2^\dagger(\vec{k'})]_-=\frac{4\omega^4}{N^2_1}\delta^3(\vec{k}-\vec{k'})$
defines  a physical commutator, while  the remaining four  modes
form the quartet to give the zero norm state. The factor of 3/2 in
$[a_3(\vec{k}),a_3^\dagger(\vec{k'})]_-$ represents the
higher-derivative nature of $(\square -m^2)^3\phi_3=0$. This
corresponds to  the first case for having a physical subspace
without the non-unitarity.

For $n=4$, we have
 \be
 [A_a, A^\dg_b]_\mp=
 \frac{8\omega^6}{N^2_1}\left(
  \begin{array}{c|cccccccc}
      &  a^\dg_1 &  a^\dg_2 & a^\dg_3 & a^\dg_4 & c^\dg_1 & c^\dg_2 & c^\dg_3 & c^\dg_4 \\ \hline
  a_1 &  0 &  0 & 0 & -i & 0 & 0 & 0 & 0 \\
  a_2 &  0 & 0 & i & 1  & 0 & 0 & 0 & 0 \\
  a_3 &  0& -i & -1 & \frac{3i}{2}  & 0 & 0 & 0 & 0    \\
  a_4 &  i &   1 & -\frac{3i}{2} & -\frac{5}{2}  & 0 & 0 & 0 & 0    \\
  c_1 &   0 & 0 & 0 & 0 & 0 & 0 & 1 & i   \\
  c_2 &   0 & 0 & 0 & 0 & 0 & 0 & -i & 0   \\
  c_3 &   0 & 0 & 0 & 0 &  -1 &  -i & 0   & 0\\
  c_4 &   0 & 0 & 0 & 0 & i & 0 & 0 & 0 \\
  \end{array}
 \right)\delta^3(\vec{k}-\vec{k'}),
 \ee
which form an octet to give the zero norm state, leaving the vacuum.
The factor of $-5/2$ in $[a_4(\vec{k}),a_4^\dagger(\vec{k'})]_-$
denotes the higher-derivative nature of $(\square -m^2)^4\phi_4=0$.

 For $n=5$, we have
 \be
 [A_a, A^\dg_b]_\mp=
 \frac{16\omega^8}{N^2_1}\left(
  \begin{array}{c|ccccccccc}
        &  a^\dg_3 &  a^\dg_1 & a^\dg_2 & a^\dg_4 & a^\dg_5 & c^\dg_1 & c^\dg_2 & c^\dg_3 & c^\dg_4 \\ \hline
  a_3 & 1 & 0 &  0 & -i & \frac{3i}{2} & 0 & 0 & 0 & 0   \\
  a_1 &   0& 0 & 0 & 0 & -i & 0 & 0 & 0 & 0  \\
  a_2 & 0& 0 & 0 & -1 & 1   & 0 & 0 & 0 & 0   \\
  a_4 & i & 0 &  -1 & \frac{3}{2} & -\frac{5}{2}  & 0 & 0 & 0 & 0    \\
  a_5 & -\frac{3i}{2} &   i & 1 & -\frac{5}{2} & \frac{35}{8}  & 0 & 0 & 0 & 0   \\
  c_1 & 0 & 0 & 0 & 0 & 0 & 0 & 0 & -i &  1 \\
  c_2 & 0 & 0 & 0 & 0 & 0 & 0 & 0 & -1 & 0 \\
  c_3 & 0 & 0 & 0 & 0 & 0 & -i & 1 & 0 &   0 \\
  c_4 & 0 & 0 & 0 & 0 & 0 & -1 & 0 & 0 & 0\\
  \end{array}
 \right)\delta^3(\vec{k}-\vec{k'}),
 \ee
which indicates that
$[a_3(\vec{k}),a_3^\dagger(\vec{k})]_-=\frac{16\omega^8}{N^2_1}\delta^3(\vec{k}-\vec{k'})$
defines a physical commutation relation, whereas the remaining eight
modes form an octet to give the zero norm state. Here we observe
that for odd $n$, there exists only one physical commutator of
$[a_2(\vec{k}), a^\dg_2(\vec{k})]_-$ for $n=3$ and $[a_3(\vec{k}),
a^\dg_3(\vec{k})]_-$ for $n=5$ which show that they describe the
states with positive norm, while for even  $n$, all commutators
belong to a quartet for $n=2$ and an octet for $n=4$ which indicate
unphysical modes. A factor 35/8 in
$[a_5(\vec{k}),a_5^\dagger(\vec{k'})]_-$ reflects  the
higher-derivative nature of $(\square -m^2)^5\phi_5=0$.

Inductively, we insist that there remains a physical subspace with
positive norm for odd $n$, but there exists  the vacuum only for
even $n$.

Finally, Ref.~\cite{Rivelles:2003jd} has stated  that all higher
derivative scalar theories are trivial after performing the BRST
quantization. However, looking  into his model equation of (1)
closely, it describes  even power of higher derivative operators
only and thus, his model hits our even $n$ case, leaving the odd $n$
untouched.

\section{General quartet mechanism and truncation mechanism}

In this section we compare the general quartet mechanism in
Minkowski space with the truncation mechanism in the AdS/CFT
correspondence.

Before we proceed, the authors~\cite{Apolo:2012vv} have discussed
the specific case of non-linear critical gravity of rank-3 in
AdS$_3$ and AdS$_4$ spacetimes with the result that truncations that
appear to be unitary at the linearized level may be inconsistent at
the non-linear level. The argument given there seems to extend to
the general case independently of how the linearized theory is
completed. This suggests that the unitary subsector  might exist
only in the linearized approximation.

Here we use the same bilinear action (\ref{action}), but the
difference is the background spacetime: the $n$ coupled scalar
theory in Minkowski and the $n$ coupled scalar theory (a toy model
of the polycritical gravity) on AdS$_3$ spacetime.

The truncation mechanism without FP ghosts   was used  to resolve
the non-unitarity in the LCFT, dual to the fourth-order  critical
gravity on AdS$_3$
spacetime~\cite{Bergshoeff:2012ev,Kleinschmidt:2012rs}.
 A rank $n$
of the LCFT refers to the dimensionality of the Jordan cell on the
boundary, while it represents the $2n$-order  polycritical gravity
on the bulk side. Explicitly, the two-point correlation functions in
the  rank-$n$ LCFT are given by
 \be \label{lcft}
 <{\cal O}^i{\cal O}^j> \sim
 \left(
  \begin{array}{ccccccc}
    0   &  & \cdots & 0       & 0     & 0      & {\rm CFT} \\
    0   &  & \cdots & 0       & 0     & {\rm CFT}    & {\rm L} \\
    0   &  & \cdots & 0       & {\rm CFT}   & {\rm L}      & {\rm L}^2\\
 \vdots &  &        & \iddots & {\rm L}     & {\rm L}^2    & {\rm L}^3  \\
        &  &        & \iddots &\iddots&\iddots & {\rm L}^4  \\
 \vdots &  &        &         &       &        & \vdots\\
  \end{array}
 \right),
 \ee
where $i,j={\rm KG},~{\rm log},~{\rm log^2},\cdots$. KG represents
the Klein-Gordon correlation function, CFT denotes the CFT
correlation function, L represents log-correlation function, and
L$^2$ is log$^2$-correlation function, etc.   For example, the LCFT
dual to a fourth-order critical gravity has a rank-2 Jordan cell and
thus, an operator has a log-mode as a logarithmic partner. For a
$2\times2$ submatrix of the top-right (\ref{lcft}), one may truncate
out  L by imposing the AdS boundary conditions to avoid the
non-unitarity.    After truncation of the rank-2 LCFT, there remains
nothing (0) for the unitary CFT.  This  is also the case for all
even rank-$n$ LCFTs.  On the other hand, the LCFT dual to a
sixth-order tricritical gravity has a rank-3 Jordan cell and an
operator has two logarithmic partners. For a $3\times3$ submatrix of
the top-right (\ref{lcft}), we throw away all correlation functions
which generate the third column and row of this matrix. Hence the
non-zero correlation functions is proportional to the unitary CFT
correlation function. Actually, a truncation may allow an odd
rank-$n$ LCFT to be a unitary CFT, while all remaining correlators
of an even rank-$n$  LCFT vanish and the theory  contains null
states after the truncation.

For the $n$ coupled scalar field theory  without the FP ghost modes
in Minkowski spacetime, the commutation relations can be recast into
the following matrix form:
 \be \label{gcrs}
 [A_i, A^\dg_j]_-=
 \frac{(-1)^{n-1}}{\alpha_1\alpha_n}\left(
  \begin{array}{c|ccccccccc}
       &  a^\dg_1 & a^\dg_2 &        &        &\cdots  & a^\dg_{n-3} & a^\dg_{n-2} & a^\dg_{n-1}   & a^\dg_n     \\ \hline
  a_1  &    0     &   0     &\cdots  & 0      &   0    &    0        &    0        &    0          &   -1         \\
  a_2  &    0     &   0     &\cdots  & 0      &   0    &    0        &    0        &   1          &   -i         \\
\vdots & \vdots   & \vdots  &\cdots  & 0      &   0    &    0        &    -1        &   i          & \frac{3}{2}    \\
       &          &         &\cdots  & 0      &   0    &   1        &    -i        & - \frac{3}{2}  & \frac{5}{2}i   \\
       &          &         & \cdots & 0      &   -1    &   i        & \frac{3}{2}& - \frac{5}{2}i &  -\frac{35}{8}   \\
       &          &         &\iddots &\iddots &\iddots &  \iddots    &   \iddots   & \frac{35}{8} &  -\frac{63}{8}i  \\
       &          &         &        &        &        &             & \iddots     &  \iddots      &   \vdots        \\
  \end{array}
 \right)\delta^3(\vec{k}-\vec{k'}),
 \ee
where $\alpha_n=\frac{(iN_1)^n}{(2\omega^2)^{n-1}}$ is the
coefficient of Eqs.~(\ref{sol1})-(\ref{sol5}), etc. Roughly, the
correlators in (\ref{lcft}) are replaced by commutators in
(\ref{gcrs}).  Also, we observe that (\ref{lcft}) and (\ref{gcrs})
have the same lower triangular matrix.   For a $2\times2$ submatrix
of the top-right (\ref{gcrs}), one may truncate out  the second
column and row  by hand to avoid the non-unitarity. After
truncation, there remains nothing (0) for a unitary scalar theory.
This is also the case for all even $n$ coupled scalar theory.  On
the other hand, for a $3\times3$ submatrix of the top-right
(\ref{gcrs}), we throw away all modes which generate the third
column and row of this matrix. Hence the only non-zero commutator is
$[a_2,a^\dg_2]_-=1$ for $\alpha_1=1/\alpha_3$. Actually, a
truncation allows an odd $n$ coupled scalar theory to be a unitary
scalar theory with positive norm states, while all commutators of an
even $n$ coupled scalar field theory  vanish and the theory contains
null states after truncation.

Hence, it is evident from the above that  without  the FP ghosts,
there is no consistent way to remove the ghost states which arise
from the higher derivative theories of $n$ coupled scalar and
polycritical gravity theories.    Noting that our action
(\ref{action}) without the FP ghosts on AdS$_3$ is dual to the
rank-$n$ LCFT~\cite{Bergshoeff:2012sc,Moon:2012wr}, it is not enough
to find a unitary  CFT consistently.    We need the $n$ coupled
scalar action (\ref{action}) as well as its FP action
(\ref{g-action}) to confine all  unphysical  fields  to  the zero
norm state, arriving at the unitary scalar theory with positive norm
states.  Finally, we insist that the truncation mechanism is nothing
but a general quartet mechanism when including  the FP ghost action.
 As was pointed out previously, the
truncation mechanism is valid for the linearized
theory~\cite{Apolo:2012vv}.

\section{Summery and Discussions}

We first summarize our main results. \\
$\bullet$ We have considered the degenerate $n$ coupled scalar field
theory (\ref{action}). For non-degenerate case with different masses
for  $W_{ij}$ in (\ref{evenm}) and (\ref{oddm}), we could not find
the BRST invariant action because the BRST symmetry is not
nilpotent. This implies that a higher-derivative Lee-Wick
model~\cite{Lee:1969fy,Carone:2008iw} is not suitable for a
consistent quantized scalar model, even though it shows clearly
which one  has positive (negative) norm states. \\
 $\bullet$ The $n=2$ corresponds to a dipole ghost field for the
 singleton~\cite{Flato:1986rx,Kogan:1999bn,Myung:1999nd}.
They form a quartet to give the zero norm state when including the
FP ghost action, leaving the vacuum only.
  \\
 $\bullet$ The $n=3$ case is enough to have a physical
subspace with positive norm states. This implies that the six-order
derivative theory~\cite{Kim:2013waf} provides a physical scalar
field.
No higher than $n=3$ coupled scalar theory is necessary to give a unitary scalar theory. \\
$\bullet$ Without the FP ghost action, we could not obtain the
consistent
truncation mechanism. This is why we have constructed the FP ghost action (\ref{g-action}) which has  non-trivial terms for $n>3$ when comparing the known cases of $n\le 3$.\\
$\bullet$ The truncation mechanism becomes the  general quartet
mechanism
 when introducing FP ghost action. The truncation mechanism works for the boundary CFT theory via the AdS/CFT correspondence,
 while the general quartet mechanism works for the bulk theory  of the $n$ coupled scalar theory in Minkowski spacetime.
 In this sense, the general quartet mechanism is dual to the truncation mechanism. \\
$\bullet$ The physical field is given by  the  $\phi_{k+1}$ for the
odd $n$ coupled scalar theory. This implies that even though
$\phi_n$ satisfies $(\square-m^2)^n\phi_n=0$  and $\phi_l$ are
regarded as auxiliary fields in the classical aspect,
$\{\phi_n,\phi_l\}$ are treated equally as scalar fields in the
quantum aspect. A centered field $\phi_{k+1}$ between $\phi_1$ and
$\phi_{n}$  is considered as a physical field with positive norm
state in the odd $n$
coupled scalar field theory.\\
 $\bullet$ We need to introduce the higher-order FP ghosts when
 quantizing the higher-order derivative gauge and gravity theory in addition to the
 gauge FP ghosts.

 Finally, we wish  to mention our implications to quantum gravity.
 Even though our model is a non-interacting scalar field model in Minkowski spacetime, the
 similar statements could be made for interacting spin-2 models. In
 the three-dimensional AdS gravity theory,  the most general
 Einstein-Hilbert action is given by~\cite{Bergshoeff:2012ev}
 \begin{equation} \label{six2}
 I_{\rm 3DAdS}=\frac{1}{16\pi G}\int d^3x\sqrt{-g} \Big[\sigma R-2\Lambda_0+\alpha R^2+\beta R^{\mu\nu}R_{\mu\nu}+{\cal L}_{\rm R\square R} \Big],
 \end{equation}
 with a sixth-derivative combination
 \begin{equation}
 {\cal L}_{\rm R\square R} = b_1\nabla_{\mu}R\nabla^{\mu}R+b_2
 \nabla_{\rho}R_{\mu\nu}\nabla^{\rho}R^{\mu\nu}.
\end{equation}
Imposing the condition of avoiding scalar gravitons,
\begin{equation}
b_1=-\frac{3}{8}b_2,~~\alpha=\frac{\Lambda}{8}b_2-\frac{3}{8}\beta,~~\Lambda=-\frac{1}{\ell^2}
\end{equation}
we finds the parity-even tricritical (PET) gravity which
 is proposed as a promising model of quantum gravity.
At the tricritical point of $\beta=-4\sigma/\Lambda$ and
$b_2=-2\sigma/\Lambda^2$, its linearized equation takes the form
\begin{equation} \label{ten-eq}
{\cal G}_{\mu\nu}({\cal G}({\cal G}(h)))=0,
\end{equation}
where the gauge-fixed linearized Einstein tensor is given by
\begin{equation}
{\cal G}_{\mu\nu}=-\frac{1}{2}(\bar{\square}-2\Lambda)h_{\mu\nu}.
\end{equation}
We note that tensor equation (\ref{ten-eq}) is similar to the $n=3$
scalar equation of $(\square-m^2)^3\phi_3=0$ when replacing
$2\Lambda $ by $m^2$. It shows that the $n=3$ coupled scalar theory
is a toy model of (\ref{six2}). Accordingly, it is possible to
reformulate the PET gravity as a two-derivative tensor theory upon
introducing  two auxiliary fields $f_{\mu\nu}$ and
$\lambda_{\mu\nu}$. This model is promising because it will be a
ghost-free theory if one introduces gauge FP ghosts for the metric
perturbation $h_{\mu\nu}$ and two higher-order FP ghosts for two
auxiliary perturbations $k_{2\mu\nu}$ and $k_{1\mu\nu}$ to lower
higher-derivative terms in the bilinear action of (\ref{six2}).
Considering the connection between $ (\phi_1,\phi_2,\phi_3)$ in the
$n=3$ coupled scalar theory and
$(k_{2\mu\nu},k_{1\mu\nu},h_{\mu\nu})$ in the PET gravity, it
conjectures that a  physical tensor would be $k_{2\mu\nu}$ with
positive norm states.
 However,  we  need
 more time to find a quantum gravity model in Minkowski spacetime~\cite{Stelle:1976gc,Barth:1983hb}
 because the tricritical gravity is still unknown in Minkowski
 spacetime.

\begin{acknowledgments}
This work was supported by the National Research Foundation of Korea
(NRF) grant funded by the Korea government (MSIP) through the Center
for Quantum Spacetime (CQUeST) of Sogang University with grant
number 2005-0049409. Y. S. Myung was also supported by the National
Research Foundation of Korea (NRF) grant funded by the Korea
government (MSIP) (No.2012-R1A1A2A10040499).

\end{acknowledgments}

 \end{document}